\documentclass[conference]{IEEEtran}
\IEEEoverridecommandlockouts
\usepackage{cite}
\usepackage{amsmath, amssymb, amsfonts}
\usepackage{algorithm, algorithmic}
\usepackage{graphicx}
\usepackage{textcomp}
\usepackage{xcolor}
\usepackage{url}
\def\BibTeX{{\rm B\kern-.05em{\sc i\kern-.025em b}\kern-.08em
		T\kern-.1667em\lower.7ex\hbox{E}\kern-.125emX}}

\begin{document}
	\title{Security Analysis on Social Media Networks via STRIDE Model \thanks{Publication with Open Access in The 19th International Conference on Networking and Services (ICNS 2023), pp. 28-33, IARIA, 2023.}}

	\author{
		\IEEEauthorblockN{Kamal Raj Sharma, Wei-Yang Chiu and Weizhi Meng}
		\IEEEauthorblockA{SPTAGE Lab, Department of Applied Mathematics and Computer Science, \\Technical University of Denmark, Denmark}
		}

	\maketitle
	\begin{abstract}
Security associated threats are often increased for online social media during a pandemic,
such as COVID-19, along with changes in a work environment. For example,
employees in many companies and organizations have started to work from
home due to the COVID-19 pandemic. Such working style has increased many
remote activities and further relied on email for communication, thus creating
an ideal condition for email fraud schemes. Motivated by this observation, the main purpose of this work is to evaluate the privacy policy of online social media and identify potential security associated problems. First, we perform a risk analysis of online
social media networks such as Facebook, Twitter and LinkedIn by using the
STRIDE model. This aims to find threats and vulnerabilities in the online social media. Then in
this analysis, the phishing attack was found to be a main threat in online social
media, which is a social engineering attack, where users are convinced through
some fake messages or emails to extract their personal credentials.
	\end{abstract}


\section{Introduction}
Social media is an Internet-based form of communication. Millions of people
around the world are using social media to share information and communicate with
each other~\cite{Ghani2019}. By using social media, people get to have conversations, share
information and create web content personally, professionally or at a company
level. There are many forms of social media popularly used currently including
blogs, micro-blogs, wikis, social networking sites, photo-sharing sites, instant
messaging, video-sharing sites, Meta-virtual worlds, Facebook, Twitter, LinkedIn,
Viber, WhatsApp and more~\cite{Meng2019}. Social networking media, especially in recent years, has
been used in different application domains, such as Government, Business, Dating,
Education, Finance, medical and health, and social and political application.
According to the Statista (German online portal for statistic), 2958, 2000, 2000, 556 million users are active users in Facebook,
Instagram, WhatsApp and Twitter, respectively, in January 2023~\cite{b1} and 734.7 million
users are active in LinkedIn by the year of 2022~\cite{b2}.

Social media can be divided into two categories: Web-based social network application and Mobile-based social network application.
	
\begin{itemize}
\item Web-based Social Network Applications:
\begin{itemize}
\item Facebook: This is a popular social networking site, alowing people to connect
with network of friends, business houses and organizations. Users can
log in using both a browser or a mobile application.

\item LinkedIn: This is a business related social media platform mainly used for
professional networking. It is an ideal site to post personal updates, job postings,
academic programs, events and projects. Users can log in using web browser.

\item Twitter: This micro-blogging site allows users to post updates. Business
houses and individuals expecting to engage with their followers at a high
frequency rate should consider using Twitter. Users can log in using both a browser
and a mobile application.

\end{itemize}

\item Mobile-based Social Network Applications.
\begin{itemize}
\item Viber: Free and secure calls and messages to anyone, anywhere. Used in
mobile application.

\item WhatsApp: Free and secure call and messages. Available in smartphones, or a web browser.

\item Telegram: A famous cloud-based instant messaging application with completely free
services.

\end{itemize}

\end{itemize}

\textbf{Motivation.} However, using too many social networking sites for conveying messages could
dilute the entire social media strategy resulting in the ineffectiveness of entire
planning and effort. So it becomes obvious that users have to be aware about
which social media sites fit into their requirements and communication strategy. For example,
it is better to choose social media sites that can be relevant to individual users.
It is also easy to connect with others in social media by making new friends, creating
new jobs or sharing new information whether it is for business or personal reasons~\cite{Fu2022}. However, there is a high risk of leaking private information and misusing the personal information because the bad actors can utilize those information for their own gain. Below are some motive examples about the risk:

\begin{itemize}
\item \textbf{Post information and update status:} Sensitive information may be
revealed. It allows users to update status anyone, anywhere at any time.

\item \textbf{Friends' Requests:} Carelessness in accepting friends'
requests may result in adding `enemies' instead of `friends' who
have more access to users' information.

\item \textbf{Upload photos and videos:} It allows everyone to view photos
and videos that are sensitive to
either a user or an organization.

\item \textbf{Third party applications and links to external sites:} While operating the applications or
clicking on the links, malware may infect users' computing platforms.
\end{itemize}

\textbf{Contributions.} Though there are many studied investigating the risk of social media networks, to our knowledge, STRIDE model has not been used to analyze most social media applications. Also, due to the spread of COVID-19, there might be a change in the security landscape.   Motivated by these, our work aims to bring in light how threats are increasing in the use of social media
and most importantly how the attackers make most use of pandemic like the
COVID-19 to spread the malicious contents through phishing attacks.
In order to reach the aforementioned conclusion, a risk analysis on social media
was performed, which results in phishing attack being the main threat in online
social media that is in increment with the ever growing use of social media.

The remaining parts are structured as follows. Section~\ref{sec:2} introduces related research on risk/security analysis on social media networks. In Section~\ref{sec:3}, we explain our security analysis outcomes based on Facebook, Twitter and LinkedIn via STRIDE Model. Section~\ref{sec:4} discusses the form of phishing attack under COVID-19 situation and provides relevant countermeasures. Section~\ref{sec:5} concludes our work.

\section{Related Work} \label{sec:2}
Risk in online social networks (OSNs) has received much attention from around 2010/2011. For instance, Tang \emph{et al.}~\cite{Tang2011} introduced an early work that identified the privacy risks due to the lack of symmetric configurations in most of the OSNs, and designed a inference attack that can be used to infer users' private information, even users already made their friend list private. Creese \emph{et al.}~\cite{Creese2012} figured out one key question about unchecked publishing and sharing of content and information in OSNs, and introduced a model to understand the potential risks faced should all of existing tools and methods be accessible to a malicious entity. The model enables easy and direct capture of the data extraction methods through the encoding of a data-reachability matrix.
	
Yang \emph{et al.}~\cite{Yang2014} then figured out that users usually group their friends into social circles but the circles are not formed with privacy policies. They introduced a utility-based trade-off framework that models users' concerns  and incentives of sharing, and made a trade-off between these two. Chan and Saqib~\cite{Chan2015} showed that online social circles such as `Facebook friends' are akin to collectivistic communities by offering users a `cushion' that mitigates financial loss, which increases users' financial risk-taking, consistent with the cushion hypothesis. Laleh \emph{et al.}~\cite{Laleh2015} introduced a risk measure, called \emph{local risk factor}, with the key idea that the malicious users in OSNs may show some common features on the topology of their social graphs, which are different from those of legitimate users.

Aktypi \emph{et al.}~\cite{Aktypi2017} examined the potential exposure of users' identity that is caused by information that they share online and personal data that are stored by their trackers. They then developed a tool to model online information shared by individuals and elaborated on how they might be exposed to the unwanted leakage. van Schaik \emph{et al.}~\cite{vanSchaik2018} focused on the security- and privacy settings of Facebook, and found there is a need for non-aggregated analysis and practical implications emphasise interventions to promote safe online social-network use. Han \emph{et al.}~\cite{Han2019} found that OSN users try to hide some information for privacy, but the hidden information is likely to be predicted by various powerful inference attacks. Then they proposed a general Framework for Private Attribute Disclosure estimation (F-PAD), which can estimate the disclosure risk for individuals in terms of disclosure probability and risk level.

Chen \emph{et al.}~\cite{Chen2020_tdsc} focused on inference attack defence, and formulated the social network data sharing problem through an optimization-based approach. Then they proposed two privacy-preserving social network data-sharing methods to counter the inference attack. One is called the efficiency-based privacy-preserving disclosure algorithm (EPPD), and the other is to convert the original problem into a multi-dimensional knapsack problem (d-KP) using greedy heuristics. Fu and Yao~\cite{Fu2022} introduced an effective and reasonable privacy risk scoring method. It takes into account the granularity of the shared profile items, combines sensitivity and visibility, and generates a privacy risk score for each user.

There are many previous studies on this topic, but to the best of our knowledge, the STRIDE model~\cite{Eldewahi2018} has not been widely used for risk analysis on OSNs. This indeed motivates our work, especially under COVID-19 situation, there could be some new attack vectors.

\section{Security Analysis with STRIDE Model} \label{sec:3}
	
The rapid increase of online social media may also bring new types of threats that
spill over from the Internet world to everyday life~\cite{Chen2020_tdsc}. For example, it has become very easy
for an intruder to exploit social media for malicious purposes, but organizations and
governments find it difficult to accurately detect, identify, predict and prevent
the malicious exploitation of social media. In this section, we aim to perform a STRIDE
model-based risk analysis on popular social media networks such as Facebook, Twitter, and LinkedIn.
	
\subsection{STRIDE Model}
The STRIDE model was designed by Praerit and Loren at Microsoft, which can be used to threat modelling of software, hardware and network systems~\cite{Eldewahi2018}. It provides a mnemonic for security threats as shown in Table~\ref{tab:1b}.

\begin{table} [t]
\small
\centering
\caption{Threats used in STRIDE Model.}
\label{tab:1b}
\begin{tabular}{|l|l|} \hline
\textbf{Threat} & \textbf{Properties}  \\ \hline \hline
Spoofing & Authenticity \\ \hline \hline
Tampering & Integrity \\ \hline \hline
Repudiation & Accountability/non-repudiation  \\ \hline \hline
Information Disclosure & Confidentiality \\ \hline \hline
Denial of Service & Availability \\ \hline
\end{tabular}
\vspace{-3mm}
\end{table}

\textbf{Spoofing} is the process of manipulating data look like it has come from different
sources. The main goal of spoofing is to cover the attacker tracks by misleading
the server using a fake address. Examples include E-mail spoofing, MAC spoofing and IP address spoofing.

\textbf{Denial of service} is an attack in which the attacker attempts to make the victim
unavailable to its legitimate users, through a temporary or indefinite interruption
of provided services.

\begin{table*} [t]
\small
\centering
\caption{Threat model of Facebook.}
\label{tab:2b}
\begin{tabular}{|l|l|l|l|} \hline
\textbf{Threat} & \textbf{Violated Property }& \textbf{Definition }& \textbf{Example}  \\ \hline \hline
Spoofing & Authentication & Pretending to be someone else & Make fake Facebook account \\ \hline \hline
Tampering & Integrity & Modify post on user's timeline & Delete/change post and message of others \\ \hline \hline
Repudiation & Non-repudiation & Claim  the real user & Multiple accounts and profile  \\ \hline \hline
Information  Disclosure & Confidentiality & Unauthorized party gain access to Info & Malicious links, e.g., phishing URL \\ \hline \hline
Denial of Service & Availability & Service unavailable to user & Overflow system, shutting down system \\ \hline
\end{tabular}
\vspace{-2mm}
\end{table*}

\begin{table*} [t]
\small
\centering
\caption{Threat model of Twitter.}
\label{tab:3b}
\begin{tabular}{|l|l|l|l|} \hline
\textbf{Threat} & \textbf{Violated Property }& \textbf{Definition }& \textbf{Example}  \\ \hline \hline
Spoofing & Authentication & Pretending to be someone else & Make fake Twitter account \\ \hline \hline
Tampering & Integrity & Retweet false news  & Promote false news \\ \hline \hline
Repudiation & Non-repudiation&  Claim the real user & Multiple accounts and profile  \\ \hline \hline
Information  Disclosure & Confidentiality & Unauthorized party gain access to Info & An unauthorized person composes
and \\
& & & sends tweets via text messages from a   \\
& & & phone number associated with account \\ \hline \hline
Denial of Service & Availability & Service unavailable to user & Overflow system, shutting down system \\ \hline
\end{tabular}
\vspace{-2mm}
\end{table*}

\textbf{Tampering} in STRIDE models means any improper modification of information.
\textbf{Repudiation} is the ability to deny the participation in the communication or
part of it. For example, the attacker can log into the system that does not have
a log or tracing program running, so there is no evidence to decide who does
what. \textbf{Non-repudiation} is to ensure that this repudiation does not occur.

\textbf{Information Disclosure} means to spying the information by attackers rather than his/her
direct intention. For example, when a web server has a crash, there will be an error message utilized by administrators to discover the problem, but it may also give the attacker a chance to compromise the server.

For the properties of Information Security, STRIDE model mainly considers the followings:

\begin{itemize}
\item \textbf{Confidentiality:} Preserving authorized restrictions on information access and
disclosure, including means for protecting personal privacy and proprietary information.

\item \textbf{Integrity:} Guarding data against improper modification or destruction of information.

\item \textbf{Availability:} Ensuring the timely and reliable use of and access to information.

\item \textbf{Authenticity:} The property of being genuine and being able to be verified
and trusted.

\item \textbf{Accountability/Non-repudiation:} The goal that generates the requirement
for actions of an entity to be traced uniquely to that entity.
\end{itemize}

Below are the assets and objects that are critical for online social media.

\begin{itemize}
\item \textbf{Hardware:} Personal computer, mobile phone, data store server, etc.

\item \textbf{Software:} Web browsing, mobile application, etc.

\item \textbf{Data:} User information at the server.
\end{itemize}

During the security analysis, a number of threats have been identified, which we
need to protect against to ensure that the security goals of the system are
achieved.


To assess and determine the risk levels of different threats, risks are modeled with probabilities
and impacts. In this work, both probability and impact are defined in three levels
(low, medium, high).

\begin{itemize}
\item \textbf{Low:} A successful attack does not affect the functionality of a system.
\item \textbf{Medium:} Requires active action, but does not render the system unable to function indefinitely.
\item \textbf{High:} Irreversible or fatal damage to the system.
\end{itemize}

We show how to categorize an
attack in different probability levels as follows:

\begin{itemize}
\item \textbf{Low:} The resources required for the attack outweigh the gain even if the
attack is successful.
\item \textbf{Medium:} The resources required for the attack are comparable to the
gain of a successful attack.
\item \textbf{High:} The gain from a successful attack should outweigh the resources needed
to perform the attack.
\end{itemize}

\begin{table*} [t]
\small
\centering
\caption{Threat model of LinkedIn.}
\label{tab:4b}
\begin{tabular}{|l|l|l|l|} \hline
\textbf{Threat} & \textbf{Violated Property }& \textbf{Definition }& \textbf{Example}  \\ \hline \hline
Spoofing & Authentication & Pretending to be someone else & Fake job offer by using fake profile \\ \hline \hline
Tampering & Integrity & Target potential victim & Convince users to open an email \\ \hline \hline
Repudiation & Non-repudiation&  Claim the real user &  Multiple accounts and profile  \\ \hline \hline
Information  Disclosure & Confidentiality & Unauthorized party gain access to Info  & Malicious links, e.g., phishing \\ \hline \hline
Denial of Service & Availability & Service unavailable to user & Overflow system, shutting down system \\ \hline
\end{tabular}
\vspace{-1mm}
\end{table*}

\subsection{Security Analysis of Facebook}
Facebook introduced a privacy policy in 2009 for the first time, where users
could select a personal privacy setting for their personal data. However, the default
option was selected to be ``Everyone", so many users accepted the default
setting without being aware of the risks. This allowed much of the data to become
publicly available.

After receiving feedback and criticisms about privacy concern, Facebook proposed a new privacy setting in 2010.
There were different levels of privacy setting options on the page including Everyone, Friends of
Friends, Friends Only for each data category. However, it was not sufficient
to prevent privacy for users. Facebook did not possess strong privacy till
2011, where people could not reach some users' personal data and profiles without
being friends.

Table~\ref{tab:2b} provides the threat model of Facebook. In most of the cases, attackers make use of the Facebook infrastructure to gather
and expose the personal information of users and their friends. In doing so
the attackers are able to make them go to malicious links, advertisements by
generating fake profiles. Some of such common attacks in Facebook are shown below:

\begin{itemize}
\item \textbf{Compromised Account Attacks:} A compromised account is the condition of an account in which legitimate
users lose complete or partial control over the login credentials~\cite{Egele2017}. Accounts can be compromised in different ways, e.g., by using a phishing scam to gather user login credentials, by utilizing cross-site
scripting, and adopting bots to harvest login credentials. Compromised
accounts can be very powerful means to spread out the malicious contents
that can deteriorate the relationship established by the legitimate user in
the past, and to communicate the malicious contents rapidly and effectively.

\item \textbf{Sybil Attacks:}
Malicious users create several fake identities, called Sybil, for influencing
their identity within a target network~\cite{Sun2014}. After that, such malicious users
send a friend request to rest of the users of that network. When one accepts
the friend request, the malicious users will forward the malware and spam.
Normally, Sybil attacks are found to be of two ways on Facebook. Attackers
generate several fake identities to create legitimate accounts to spread
malware and spam to friends in their friend list or form more social links
to distribute malware and spam.

\item \textbf{Socware Attacks:}
For such attack type, contenders create malware, also known as \emph{socware}, in the
form of events, applications or pages capable of having links to malicious
contents. False gift vouchers, coupons, and gifts are used as stimulants to
attract victims, and then cheat them into installing or accepting the malware~\cite{Coban2020}.
Once the malware has been installed in the system, attackers can easily gather
personal information stored by the users. On top of that, the malware is
posted on the user's wall, which will also spread on their friend's profile.

\item \textbf{Identity Clone Attacks:}
Malicious users, sometimes, create similar profiles pretending to be the
victims and outspread malicious content into their network. To make it possible, the malicious users normally first attempt to get a victim's personal information, such as occupation, name, and friends list. After collecting the
information, attackers can copy the victim's profile and sends friend request to
the victim's friends

\item \textbf{Creepers Attacks:}
Creepers are actual users who use online social media network functionalities
in a wrong way~\cite{Coban2020}. For example, they would send a friend request
to many unknown users and post spammy letters on their walls. Such
attacks are mainly used for advertisements.

\item \textbf{Cyberbullying attacks:}
Cyberbullying is one of the most known and popular attacks in social network.
This type of attacks can harass victims by posting
sexual remarks, threat, repeated hurtful messages and irrelevant and disgusting
contents. Besides, they can plant rumours about victims by posting
awkward and embarrassing videos and / or photos online. According to a
research study~\cite{Bunga2019}, 12\% of the parents complained that their children
have been cyber bullied.

\item \textbf{Clickjacking Attacks:}
For such attack type, it is also known as user interface (UI) redressing, where an adversary will trick users to click on some actionable contents, which are actually different from what they intend to click on. Afterwards, they can collect the personal
information from these users and send spam messages and malicious links
on their wall~\cite{Rehman2013}.

\end{itemize}

\subsection{Security Analysis of Twitter}

The popularity of Twitter has changed the way that
users interact with technology. Generally, the users share their data with social network sites
in a transparent way. Twitter is one of the famous public platforms, which
provides developers with Application Programming Interface but with
limited use for multiple reasons such as data volume, user privacy expectation,
and Twitter business interests, since the platform will share some private information
with advertisers such as how users interact with ads and which ones attract their
attention. Sharing data is very crucial for the Twitter company because it has
been proved that Twitter users interact with ads that advertisement companies
post. As a result, these companies will pay Twitter and help it operate a
free service.

Table~\ref{tab:3b} shows the threat model of Twitter.
It is found that Twitter may suffer from various attacks but can also be used to intrude many users.

\begin{itemize}
\item \textbf{Short-URLs:}
Due to the strict limitation on tweet length, users will use short-Universal Resource
Locators (URLs) in tweets instead of standard URLs. The short URLs are indeed
ordinary URLs that are encoded into URLs with the least characters, which
thus best suit in tweets~\cite{Gupta2014,Wang2013}. A normal user has very limited knowledge
of the target of the short URLs, and such users can easily be exploited
and manipulated to download and / or spread
malicious software without their knowledge. We have been experiencing
plenty of such attacks in the recent times, as short URLs have increased
in number along with the explosion of short messaging for mobile users.
Attackers are able to exploit human shortcomings in various ways with
the use of shortened URLs. Making it more difficult to understand and
analyse, busy users do not take time to look into the
link of the short-versioned URLs, therefore,  the underlying URLs are more probable to be
clicked. As many phishing emails are targeted to elicit quick emotional response from the recipients warning on negative consequences, an exhausted
employee may hastily click on such links. Shortened URLs also
benefit from the fact that several employees may not normally be aware
of how a shortened URL looks like.

\item \textbf{Compromise and control a user account:}
In addition to sending direct messages to the users, attackers can use a compromised
account to tweet to the followers. The probability of followers
and the other users linked to the followers clicking on such ill-motivated
links, in this case, is greater than the case of tweeting to the direct user, due to the fact that there is already a significant degree of trust between the
users and their followers.

\item \textbf{Clickjacking:}
Clickjacking method is a very common and widely implemented attack among the advanced self-propagating
attacks. The chance of clicking on a link is more likely by a follower of
a user than by any other non-followers. In such attack, tweet
retweets itself whenever a user of Twitter clicks on the link.

\item \textbf{Indirect attack:}
The clickjacking attack is remodelled to travel beyond Twitter. When
the Twitter users are surfing in other public websites that allow users
to enter links to other websites such as news sites or blogs. These sites
provide a malicious short-URLs, and clicking on the links would result in
a clickjacking attack when such a victim also has a Twitter handle.

\end{itemize}

\subsection{Security Analysis of LinkedIn}

Privacy is a great concern for LinkedIn, and that is why they have stressed upon it many times. Their main aim is to make transparency about the data they are collecting from the users. The privacy policy is applied on the users, who are using their service or product. LinkedIn provides their users an option to make a choice about the data collection, use and sharing as described in the privacy policy.
The data collected by LinkedIn starts from creating a profile or an account,
which includes user name, email address or mobile phone number and a password.
If users would like to have a premium service, then they have to provide
payment and billing information. After the registration phase, the user moves
into profile setting. A user can fill in the information regarding his / her education,
experience, skills and profile summary.

Table~\ref{tab:4b} shows the threat model of LinkedIn. It is the same that many attacks are threatening the security and privacy of LinkedIn users.

\begin{itemize}
\item \textbf{Illegitimate Contact Requests:}
Similar to other online social media platforms, the act of connecting with another
LinkedIn user also leaves enough space for malicious activities. As a
matter of fact, one of the most common tactics on LinkedIn is when a
user gets a fake connection request from another member. Such requests
may take on one of several different forms. In many cases, scammers may mostly
claim that they are romantically interested at the recipients.

\item \textbf{Fake Job Offers:}
Users, sometimes, receive a LinkedIn message from claiming to be a job recruiter. The spammer then details a high-paying job and convinces the
users that the duties can be performed from anywhere with Internet access.
Such an offer lures a number of users as it sounds too good to be true.

\item \textbf{Phishing:}
The most customary type of phishing scam involves convincing people
into opening emails or clicking on a link / url that appears to have been sent
from a legitimate business or creditable source. LinkedIn hook has been
found to have been used by more than half of social media phishing emails.
LinkedIn has become the most trusted medium to target potential victims
with more than half of all social media phishing emails using the Microsoftowned
platform as a hook. KnowBe4's tests~\cite{b3} revealed that LinkedIn
had been used in 56\% of the top phishing emails more than all other
combined social media networks. The way such scam works is that one
receives an email from someone that they might not know in person but
is a business associate that they are connected to through LinkedIn. This
kind of email would, at the first glance, look rather innocuous. Such emails
use professional language, and one would be asked to click on a link that
would direct them to a website, and the URL being used here would seem
more or less legitimate, thereby, making one even less suspicious in the
whole phishing process.
\end{itemize}

\section{Discussion on Phishing} \label{sec:4}
According to the Infosecurity Magazine~\cite{b4}, email phishing attacks have spiked by over 600\% since the end of February 2020, as many organisations and companies started working from home because of the COVID-19 restrictions. This working environment increased remote activities and the reliance on email for communication, thus, creating perfect conditions for email fraud schemes.

Criminals are taking advantage of the COVID-19 pandemic to launch phishing attacks. Below are some typical ways:

\begin{itemize}
\item \textbf{Zoom Users Become Targets:}
This is an emerging type of phishing attack, in which intruder will
send fake Zoom video-conferencing meeting notifications. This attack
is designed to steal usernames and passwords from victims' Zoom accounts. Phishers can use these credentials to log into corporate video
conferencing accounts, and try to collect passwords afterwards.

\item \textbf{To use Covid-19 in business Email for Compromised Account Attacks:}
In such attack, the phisher will send an email to the person who has access to finance information
of the organization. The phisher pretended to be a real supplier and requested for
past due invoices including the information due to the Covid-19 in new
account.

\item \textbf{Fake registration process:}
Victim may get an instant message with a link to claim for an official registration for the immediate
withdraw of money from the Government. Thus, the phisher collects a victim's
information via fake registration process.
\end{itemize}

\textbf{Countermeasures.} Phishing prevention can be reached by providing an extra layer of security in the login
process. The extra layer could be a two-factor authentication scheme, which
is a process to confirm the user's identity before he or she is granted
to access an account.

For example, two-factor authentication can be performed via Short Message Service (SMS). When the user enters username
and password, a verification code will be delivered to the other device. User can be
granted access for login when he or she enters the verification code successfully.
This method has been widely adopted in current market.
However, this solution is not very usable in some cases, i.e., it requires an
extra device from the user and causes extra cost to implement.

User awareness can help in educating the users by which they are able to
identify phishing attempts. The big success of phishing attacks is mostly due to the negligence
of users. To help reduce the phishing attacks, the awareness campaign program is very useful and important. Currently, there is no sufficient education and awareness campaign against phishing attacks. There are some anti-phishing
methodologies such as games and security awareness tools in
the server to familiarize threats like phishing attacks and identify malicious
URLs and other phishing scams. The embedded server training program
technique is used to teach users by sending a mock phishing email and
asking them to open the attached emails or URL. Once the user opens the
phishing contained email, they will find that the email contains fake information.
In this way, the mock phishing awareness campaign increases the
end user's knowledge to protect against phishing attacks.

\section{Conclusion} \label{sec:5}
Cyber criminals are now taking advantage of social media networks to collect personal
information because users make a lot of private information
public such as location, job, email, contact number and more. In this work, we provided a security analysis using STRIDE model on three major social media networks, such as Facebook, Twitter and LinkedIn. We found that phishing attack has a high probability of risk in online social media or using an online social media for launching attacks as compared with other potential attacks. Especially, criminals have taken advantage of the COVID-19 pandemic to design particular phishing attacks, i.e., the information about COVID-19 is included for
convincing the user to click on designed URL. In the end, we also provided and discussed potential countermeasures to identify phishing content. This work aims to highlight the risk and issues in current social media networks and stimulate more defensive studies.

	

\section*{Acknowledgment}
This research work was funded by the European Union H2020 DataVaults project with GA Number 871755. 

\end{document}